# Quasiparticle number fluctuations in superconductors

C.M. Wilson and D.E. Prober*

*Yale University, P.O. Box 208284, New Haven, CT 06520-8284*

We present a general theory of quasiparticle number fluctuations in superconductors. The theory uses the master equation formalism. First, we develop the theory for a single occupation variable. Although this simple system is insufficient to describe fluctuations in a physical superconductor, it is illustrative, allowing this discussion to serve as a self-contained introduction. We go on to develop a multivariate theory that allows for an arbitrary number of levels with transitions of arbitrary size between levels. We specialize the multivariate theory for two particular cases. First, we consider intrinsic quasiparticle fluctuations. In a previous Letter, these results were used to describe time-resolved measurements of thermodynamic fluctuations in a superconducting Al box [C.M. Wilson, L. Frunzio and D.E. Prober, Phys. Rev. Lett. **87**, 067004 (2001)]. Finally, we extend these results to include fluctuations due to extrinsic loss processes.

## I. INTRODUCTION

Superconductivity is a rich, physical phenomenon with many aspects that have been studied for their possible technological importance. The most basic property of superconductors, their ability to transport electrical currents without resistance, has been applied broadly for many years. A new generation of superconducting electronic devices aims to take advantage of more subtle aspects of superconductivity, including flux quantization, quantum tunneling and the quantum coherence of the superconducting state. Examples include SQUIDs, high-speed electronics[1], superconducting detectors[2] and various implementations of quantum bits for quantum information processing.[3] The ultimate sensitivity and usefulness of these devices will be determined in part by the physical processes that add noise to them.

In this article, we present a theory of one such noise source: fluctuations in the number of quasiparticle excitations. In its ground state, all of the conduction electrons in a superconductor form bound pairs, called Cooper pairs. The binding energy of the pairs is the spectroscopic gap $E_g = 2\Delta$, where $\Delta$ is the energy gap for a single excitation. At finite temperature, some pairs will be broken, resulting in single-particle excitations known as quasiparticles. In equilibrium, the average number of quasiparticles, $N^0$, is determined by thermodynamics. In particular, the average occupation of quasiparticles levels is determined by the Fermi-Dirac distribution, with the energy measured from the Fermi energy, $\varepsilon_F$, and the minimum quasiparticle energy being $\Delta$. At a microscopic level, it is the balance of quasiparticle generation and recombination that determines the average number of quasiparticles. Quasiparticle generation refers to the creation of two quasiparticle excitations when a Cooper pair is broken by a thermal phonon. Quasiparticle recombination refers to the annihilation of two quasiparticles as they form a Cooper pair and emit a phonon. Generation and recombination are random processes, meaning that individual generation or recombination events occur at random intervals.

Because of this, the instantaneous density of quasiparticles fluctuates in time. Statistical mechanics tells us that the RMS magnitude of the fluctuations is $(N^0)^{1/2}$ (Ref. 4). In a recent Letter, we confirmed the prediction for the magnitude and we demonstrated that, at low temperatures, the time scale of the fluctuations is the recombination time $\tau_R$.[5]

In this paper, we present a general theory of quasiparticle fluctuations in superconductors. (Previous treatments of quasiparticle fluctuations were restricted to basic thermodynamic arguments.[6]) We use the master equation formalism, which has been used to describe fluctuations in semiconductors for many years. The master equation formalism reproduces and expands the thermodynamic results, while also being applicable to nonequilibrium systems. In section II, we develop the theory for a single occupation variable. This simple system is insufficient to describe fluctuations in a physical superconductor, but we have included it because it is illustrative, allowing the article to serve as a self-contained introduction to researchers in superconductivity that are unfamiliar with the semiconductor research. In addition, the more complete theory will show that the simpler results of the one-variable system can be used with the appropriate definition of effective parameters. In section III, we develop a multivariate theory that allows for an arbitrary number of levels with transitions of arbitrary size between levels. We then specialize the multivariate theory to two particular cases. In section III.B, we consider intrinsic quasiparticle fluctuations where quasiparticles are only created (annihilated) in pairs due to thermal generation (recombination). In a previous Letter, this specialized case was used to describe time-resolved measurements of thermodynamic fluctuations in a superconducting Al box.[5] In section III.C, we also allow quasiparticles to be lost and created individually. This second case can be applied to systems with normal metal traps, diffusive loss, etc..

## II. SINGLE-VARIABLE MASTER EQUATION

To treat fluctuations in our system, we construct a master equation similar to the Fokker-Planck equation. This differential equation describes the probability distribution of the occupancies of various subsystems (levels). We follow the treatment by van Vliet of generation-recombination noise in semiconductors,[7] except that we generalize the description to allow for transitions that involve an arbitrary number of particles, e.g., two quasiparticles recombining. The master equation formalism can in fact predict the fluctuation of an arbitrary number of coupled levels. However, that development is not particularly illuminating. For this reason, we will start with the derivation for a two-level system described by a one-variable master equation.

We can consider one level of our system to be quasiparticles. The second level could be Cooper pairs or quasiparticles in traps or something else, depending on the exact nature of the system that we are trying to model. (In this section, we will refer to any processes that creates (annihilates) quasiparticles as a generation (recombination) process, although in general these terms have the specific meanings defined in section I. Regardless of exactly what the second level is, it is not in general independent of the first level because the total number of excitations in the two levels is constrained. For instance, the number of quasiparticles plus Cooper pairs is constrained by the total number of electrons, due to overall charge neutrality. Furthermore the creation of two quasiparticles implies the loss of one pair, and vica versa. Therefore, we only need to count the number of quasiparticles, N, and can describe our system with a one variable master equation:

$$\frac{\partial P(N,t|k,0)}{\partial t} = -[g(N)+r(N)] \cdot P(N,t|k,0) + g(N-\delta N) \cdot P(N-\delta N,t|k,0) \\ + r(N+\delta N) \cdot P(N+\delta N,t|k,0) \quad (1)$$

where $P(N,t|k,0)$ is the probability that there are N quasiparticles at time t given that there were k quasiparticles at t=0. The function g(N) is the probability per unit time that there will be a generation event in the box when there are N quasiparticles. In other words, g(N)dt is the probability of a generation event in the time interval dt. Similarly, the function r(N) describes the probability per unit time of recombination. The parameter $\delta N$ is the number of quasiparticles added (removed) by a generation (recombination) event. We can understand the structure of the master equation quite simply. It describes the rate of change of the probability that there are N quasiparticles in the system. The rate of decrease in the probability equals the probability that there are N quasiparticles times the probability per unit time that there will be a generation or recombination event. This is what the first term in the master equation represents. The rate of increase in the probability is equal to the probability that the system is one generation event away from having N quasiparticles times the probability per unit time that there will be a generation event, plus a similar term for recombination.

The master equation is a countably infinite set of coupled differential equations. Luckily, we do not need to solve the master equation for it to be useful. We can instead use the master equation to construct much simpler equations for quantities like the variance and correlation function of the fluctuations.

We begin by calculating the variance of the fluctuations. The variance is a steady-state property, so we can set the left side of the master equation to zero. If we then multiply the equation by N, and sum over all possible N, we get the simple relationship:

$$\langle g(N) \rangle = \langle r(N) \rangle$$

where the angle brackets mean the expectation value over all N. If we expand both g(N) and r(N) in a Taylor expansion in N around the equilibrium value $N^0$, we get

$$g(N^0) + \tfrac{1}{2} g''(N^0) \langle \Delta N^2 \rangle = r(N^0) + \tfrac{1}{2} r''(N^0) \langle \Delta N^2 \rangle \tag{2}$$

where the primes indicate the derivative with respect to N and $\Delta N = N - N^0$. The first order terms vanish because $\langle \Delta N \rangle = 0$ in equilibrium. In addition, in most cases $g(N), r(N) \propto N^2$ and $\langle \Delta N^2 \rangle \propto N$, so we can neglect the second order terms and simplify to

$$g(N^0) \approx r(N^0).$$

This is the reasonable statement that the generation and recombination rates must balance in equilibrium.

If we again set the left-hand side of the master equation (1) to zero, multiply by $N^2$ and sum over all N, we get the relationship

$$\left\langle \left( N + \frac{\delta N}{2} \right) \cdot g(N) \right\rangle = \left\langle \left( N - \frac{\delta N}{2} \right) \cdot r(N) \right\rangle.$$

If we again expand g(N) and r(N) around $N^0$ and use (2) to simplify, we can find the following expression for the variance of the fluctuations

$$\langle \Delta N^2 \rangle = \delta N \frac{r(N^0)}{r'(N^0) - g'(N^0)} \tag{3}$$

where we have again neglected second order terms in the final expression.

We can also use the master equation to calculate the power spectrum of the fluctuations. To do this, we first calculate the autocorrelation function of the fluctuations and then compute its Fourier transform. The autocorrelation function at lag u is defined as:

$$\Phi(u) = \langle N(0)N(u) \rangle = \sum_k \sum_j k \cdot j \cdot P(k,0;j,u)$$

where $P(k,0;j,u)$ is the *joint* probability that there are k quasiparticles at t=0 and that there are j quasiparticles at t=u. (By lag we mean the amount of time that one signal is shifted with respect to the other). We can simplify this expression by factoring the joint probability distribution into $P(k,0;j,u) = P(j,u|k,0) \cdot P(k,0)$ giving

$$\Phi(u) = \sum_k k \cdot P(k,0) \sum_j j \cdot P(j,u|k,0) = \sum_k k \cdot \langle N \rangle_k \cdot P(k,0) \qquad (4)$$

where $P(j,u|k,0)$ is the conditional probability of having j quasiparticles at t=u given that there were k at t=0 and $\langle N \rangle_k$ is the expectation value of N given that there were k quasiparticles at t=0.

To further simplify this expression, we start by deriving a differential equation for $\langle N \rangle_k$ using the master equation. In this case, we need to use the full master equation (1) without setting the time derivative equal to zero. If we multiply both sides by N and sum over all N, we get the equation

$$\frac{d}{du} \langle N \rangle_k = \delta N \big( \langle g(N) \rangle - \langle r(N) \rangle \big).$$

We cannot solve this equation explicitly, because we do not know the expectation values on the right-hand side. However, we can find an approximate solution by again expanding g(N) and r(N) around $N^0$. We find the simple result

$$\frac{d}{du}\langle \Delta N\rangle_{k-N^0} = -\frac{\langle \Delta N\rangle_{k-N^0}}{\tau} \quad ; \quad \tau \equiv \frac{1}{\delta N}\frac{1}{r'(N^0)-g'(N^0)} \quad (5)$$

where $\tau$ appears as the effective relaxation time of the fluctuations. This equation has the simple solution

$$\langle \Delta N\rangle_{k-N^0} = (k-N^0)\exp\left(-\frac{u}{\tau}\right).$$

Inserting this solution into (4) we find the autocorrelation function of the fluctuations to be

$$\Delta\Phi(u) = \langle \Delta N(0)\Delta N(u)\rangle = \langle \Delta N^2\rangle \exp\left(-\frac{u}{\tau}\right)$$

where $\langle \Delta N^2\rangle$ is the variance of the fluctuations. We can then directly compute the power spectrum, $G(\omega)$, of the fluctuations as the Fourier transform of the autocorrelation function. We find

$$G(\omega) = \frac{4\langle \Delta N^2\rangle \tau}{1+\omega^2\tau^2}.$$

We now have general expressions for the variance and power spectrum of the fluctuations in our two-level system. Before we specialize the equations more, we can make some general comments. First, if we combine (3) with (5), we find the much simpler expression for the variance of the fluctuations:

$$\langle \Delta N^2\rangle = (\delta N)^2 r(N^0)\,\tau.$$

This says that the variance of N is of order the number of particles that recombine in one correlation time. Now, looking at (5), we see that τ is inversely proportional to δN. This says the more quasiparticles that are lost (created) by a single recombination (generation) event, the faster the fluctuations. Also, looking at (5) we see that the time scale of fluctuations is inversely proportional the sum of the derivatives of the generation and recombination rates. This has a simple physical interpretation. In Fig. 1 we sketch the recombination parameter, r(N), and generation parameter, g(N), as a function of N. First, we note that the value of N where the curves intersect is the equilibrium value $N^0$. Next, we notice that for a stable system the derivative of r(N) will always be positive and the derivative of g(N) will always be negative. This is what maintains equilibrium. For example, if N fluctuates greater than $N^0$, then the recombination rate increases *and* the generation rate decreases. Both of these changes drive the system back to equilibrium. Even more, the steeper the change in the rates around equilibrium, the faster the system is driven back to equilibrium. This is why the time constants depend on the derivatives of r(N) and g(N) and why their contributions sum together.

To be able to apply the formulas derived above we must know what r(N) and g(N) are for our system. Luckily, if we already understand the dynamics of the system, it is general easy to deduce r(N) and g(N). In general, the rate equation of our system will be of the form

$$\frac{dN}{dt} = \delta N(g(N) - r(N)). \tag{6}$$

If we can derive or know an appropriate rate equation for our system, we can then read off g(N) and r(N).

We can consider, as an example, the case of simple generation and recombination of quasiparticles. By simple, we mean that quasiparticles are only lost to recombination with other quasiparticles and we ignore the effects of phonon trapping (which we will return to later). In this case, the two levels of our system are quasiparticles and Cooper pairs, with the total number of electrons constrained to be the normal state value. We will further assume that we are working at low temperatures and that the number of quasiparticles is small compared to the number of Cooper pairs. In general, we would expect g(N) to depend on the number of Cooper pairs. However, since the relative size of the fluctuations will be small compared to the number of Cooper pairs, we will assume g(N) is constant and equal to the equilibrium recombination rate. With that we can write the rate equation for our simple system as

$$\frac{dN}{dt} = 2\left(\Gamma_G - \frac{1}{2}\frac{R}{Vol}N^2\right)$$

where $\Gamma_G$ is the constant generation rate, *Vol* is the volume of the system, and R is the recombination constant. The recombination constant is basically a constant of proportionality between the recombination rate and the number of ways to combine N quasiparticles, which is $N^2/2$.

From this rate equation, we can read off the parameters of our model:

$$g(N) = \Gamma_G \quad ; \quad r(N) = \frac{1}{2}\frac{R}{Vol}N^2 \quad ; \quad \delta N = 2.$$

We can then easily put these parameters into the equation above to find a familiar result for the variance of the fluctuations, $\langle \Delta N^2 \rangle = N^0$. We can also easily write down the power spectrum of the fluctuations

$$G(\omega) = \frac{4N^0 \tau}{1+\omega^2 \tau^2} \quad ; \quad \tau = \frac{Vol}{2RN^0}. \tag{7}$$

We see that the spectrum has a simple Lorentzian form with a bandwidth given by $1/\tau$.

## II. MULTIVARIABLE MASTER EQUATION

### A. General theory

The simple one-variable master equation derived above is illustrative, but it is not sufficient to describe generation and recombination in a physical superconductor. For example, in a thin-film superconductor the phonon emitted when a pair of quasiparticles recombines can break another pair before the phonon escapes the film into the bath. This process, known as phonon trapping, extends the effective lifetime of a quasiparticle. To account for this process, or others like it, we must increase the number of levels in our model system. We can treat the fluctuations of a multilevel system with a multivariable master equation. The basic idea is the same as before, except we now describe the state of the system with levels (1) - (S) by a vector $\mathbf{a} = (N_1, N_2, ..., N_S)$ which represents the occupation of each level. In general only S-1 levels will be independent since the total number of excitations is constrained. We start by writing down the master equation for the process:

$$\frac{\partial P(\mathbf{a},t|\mathbf{a}',0)}{\partial t} = \sum_{\mathbf{a}''\neq\mathbf{a}} P(\mathbf{a}'',t|\mathbf{a}',0) Q(\mathbf{a};\mathbf{a}'') - \sum_{\mathbf{a}''\neq\mathbf{a}} P(\mathbf{a},t|\mathbf{a}',0) Q(\mathbf{a}'';\mathbf{a}) \tag{8}$$

where $P(\mathbf{a},t|\mathbf{a}',0)$ is the probability that the system is in state $\mathbf{a}$ at time t given that it was in state **a'** at t=0, etc. and $Q(\mathbf{a};\mathbf{a}'')$ is the transition probability per unit time from state **a''** to **a**. Again, the first term says that the rate of change in the probability of finding the system in state **a** is the probability of it being one transition away from **a** times the rate of transition to **a**. The second term accounts for transitions out of state **a**. We can make this less abstract if we notice that the only allowed transitions in our system involve a single loss event in one level causing a creation event in a second level. We can then write

$$Q(\mathbf{a};\mathbf{a}'') = \begin{cases} p_{ij} \; ; \; \mathbf{a}'' = \{n_1, \ldots, n_i, \ldots, n_j, \ldots\} \\ \quad\quad \mathbf{a} = \{n_1, \ldots, n_i - \delta n_{ij}, \ldots, n_j + \delta n_{ji}, \ldots\} \\ 0 \; ; \; \text{otherwise} \end{cases}$$

where $\delta n_{ij}$ is the "shot size." The physical meaning of $\delta n_{ij}$ is the change in the occupation of level $i$ when making a transition to or from level $j$. This is one important generalization of the master equation formalism for superconductors. In typical semiconductor systems, transitions between all levels change the occupation by one, i.e., $\delta n_{ij} = \delta n = 1$ for all transitions. In superconductors, however, not only can different levels have a different shot size, they can have a different shot size depending on what the other level involved in the transition is.

We can then proceed along the same lines as the derivation in section **I**. We will not include the detailed derivation, instead presenting the results and referring to Ref. 7 for a more detailed treatment. In analogy to the linearized time constant found in (5), we can write a linearized rate matrix, **M**, where the elements are:

$$M_{ij} = \sum_k \delta n_{ik} \left( \frac{\partial p_{ik}}{\partial N_j} - \frac{\partial p_{ki}}{\partial N_j} \right)\bigg|_{\{N_i\}=\{N_i^0\}}. \tag{9}$$

We can define a second matrix, **B** (which describes the second order Fokker-Plank moments), whose elements are:

$$B_{ii} = \sum_{k \neq i} \delta n_{ik}^2 (p_{ki} + p_{ik}) \approx 2 \sum_{k \neq i} \delta n_{ik}^2 p_{ik}^0$$
$$B_{ij} = -\delta n_{ij} \delta n_{ji} (p_{ij} + p_{ji}) = -\delta n_{ij} \delta n_{ji} (p_{ij}^0 + p_{ji}^0)$$
(10)

The covariance matrix, $\boldsymbol{\sigma}^2 = \langle \Delta \mathbf{a} \Delta \mathbf{a}^T \rangle$, is then determined by the following matrix equation

$$\boldsymbol{\sigma}^2 \mathbf{M}^T + \mathbf{M} \boldsymbol{\sigma}^2 = \mathbf{B} \qquad (11')$$

where $\Delta \mathbf{a} = \mathbf{a} - \mathbf{a}^0$. We can also write the cross power spectrum matrix as

$$\mathbf{G}(\omega) = 2 \operatorname{Re}\left[ (\mathbf{M} + i\omega \mathbf{1})^{-1} \mathbf{B} (\mathbf{M}^T - i\omega \mathbf{1})^{-1} \right] \qquad (12')$$

where Re[ ] means the real part and **1** is the identity matrix. The diagonal terms of **G** describe the power spectrum of the fluctuations of each level in the system. The off-diagonal terms of **G** describe the cross power spectrum between the various levels. Each spectrum $G_{ij}$ is a sum of individual Lorentzian spectra, like (7), with characteristic frequencies determined by the eigenvalues of **M**.

Eqn. (11')-(12') can be simplified for systems that are in (quasi-) equilibrium. Specifically, they can be simplified in systems that have a symmetric correlation matrix, i.e., systems where

$$\langle \Delta \mathbf{a}(t) \Delta \mathbf{a}^T(0) \rangle = \langle \Delta \mathbf{a}(0) \Delta \mathbf{a}^T(t) \rangle.$$

If this condition holds, then we can demonstrate that $\boldsymbol{\sigma}^2 \mathbf{M}^T = \mathbf{M} \boldsymbol{\sigma}^2$, and (11') and (12') reduce to

$$\boldsymbol{\sigma}^2 = \langle \Delta \mathbf{a} \Delta \mathbf{a}^T \rangle = \tfrac{1}{2} \mathbf{M}^{-1} \mathbf{B} \qquad (11)$$

and

$$\mathbf{G}(\omega) = \frac{2}{\omega^2} \operatorname{Re}\left[ \left( \mathbf{1} + \frac{\mathbf{M}}{i\omega} \right)^{-1} \mathbf{B} \right] \qquad (12)$$

In an equilibrium system, these simplified results always apply and are a consequence of microscopic reversibility.[8] They also approximately apply to systems in quasiequilibrium, which we define as a steady-state condition that obeys the principle of detailed balance, i.e., where $p_{ij}^0 = p_{ji}^0$ for all $i$ and $j$ (Ref. 9). Detailed balance always applies in equilibrium, but it can also be true in nonequilibrium steady-state, depending on the details of the level structure. In particular, it can apply in steady-state systems where levels are coupled in pairs.

An important special case is the quasiequilibrium, two-variable result. Since we are always free to label the quasiparticles as level 1, we will give the general expression for $G_{11}$ in the two variable case:

$$G_{11}(\omega) = 2\sum_{1,2} \frac{\tau_1 \tau_2}{\tau_2 - \tau_1} \frac{\tau_1^2}{1+(\omega\tau_1)^2} \left[ \left(\frac{1}{\tau_1} - M_{22}\right) B_{11} + M_{12} B_{12} \right] \quad (13)$$

where $\gamma_i = 1/\tau_i$ are the eigenvalues of **M** and the summation means add another term with $\tau_1$ and $\tau_2$ interchanged. The result for $G_{22}$ has the same form but with the indices 1 and 2 interchanged on the components of **M** and **B**.

*B. Intrinsic Quasiparticle Fluctuations*

The first specific example that we will consider is intrinsic quasiparticle fluctuations in a thin-film superconductor. By intrinsic fluctuations we mean: 1) that quasiparticles are only created in pairs through generation, whereby a Cooper pair is broken by a high-energy phonon and 2) quasiparticles are only lost in pairs through recombination, whereby a Cooper pair is formed with the emission of a high-energy phonon. This system can be described by three levels whose populations are labeled by N, $N_\omega$, and $N_{\omega,B}$ which are the number of quasiparticles in the electrodes, the number of phonons with energy $E_\omega > 2\Delta$ in the electrodes, and the number of phonons with $E_\omega > 2\Delta$ in the bath respectively. We only keep track of phonons with $E_\omega > 2\Delta$ because they are the only phonons that can generate new quasiparticles.

In section II, we thought of two quasiparticles recombining to form a Cooper pair, instead of quasiparticles recombining to form a phonon. In the end, however, $N_\omega$ is a more natural variable than the number of Cooper pairs for several reasons. From a statistical point of view, we can account for the recombination of two quasiparticles equally well as a transition to a Cooper pair or a transition to a phonon. From a dynamical point of view, however, keeping track of phonons is much more important then keeping track of Cooper pairs. As we will see shortly, the presence of phonons created by recombination can significantly change the effective recombination rate measured in experiments. On the other hand, the rate $\Gamma_B$ at which phonons break pairs and generate quasiparticles *is* proportional to the number of Cooper pairs, but as long as the number of pairs is much greater than the number of quasiparticles, then $\Gamma_B$ is approximately constant. Thus, we see that $N_\omega$ is a better choice.

We can describe the dynamics of the levels with the following system of three coupled differential equations:

$$\frac{dN}{dt} = 2\left\{-\frac{1}{2}\frac{RN^2}{Vol} + \Gamma_B N_\omega\right\} \tag{14}$$

$$\frac{dN_\omega}{dt} = \frac{1}{2}\frac{RN^2}{Vol} - \Gamma_B N_\omega - \Gamma_{ES} N_\omega + \Gamma_K N_{\omega,B} \tag{15}$$

$$\frac{dN_{\omega,B}}{dt} = \Gamma_{ES} N_\omega - \Gamma_K N_{\omega,B} \qquad (16)$$

where $\Gamma_{ES}$ is the rate at which phonons escape from the electrode to the bath and $\Gamma_K$ is the rate at which phonons enter the electrode. The factor of 2 in the first equation comes from the fact that quasiparticles are generated and recombine in pairs. We have neglected the anharmonic decay of phonons as a loss process because it happens on a time scale much longer than phonon escape at these energies.

We can simplify these equations with the approximation that $N_{\omega,B}$ is constant, which is justified because the exchange of phonons with the junction is a very small perturbation to the bath. This simplification reduces (16) to the equality $\Gamma_B N_\omega^0 = \Gamma_K N_{\omega,B}^0$, where the superscripts indicate steady-state values. We can then rewrite (15) as:

$$\frac{dN_\omega}{dt} = \frac{1}{2} \frac{RN^2}{Vol} - \Gamma_B N_\omega - \Gamma_{ES}(N_\omega - N_\omega^0) \qquad (15')$$

We then see that (14) and (15') are the well known Rothwarf-Taylor equations.[10]

Following Gray,[11] we can linearize these equations for small perturbations by writing $N = N^0 + \Delta N$ and $N_\omega = N_\omega^0 + \Delta N_\omega$ and simplifying. If we define the vector $\mathbf{a} = (N, N_\omega)$ then we can write the linearized equations in matrix form

$$\frac{d(\Delta \mathbf{a})}{dt} = -\Gamma \cdot \Delta \mathbf{a} \quad ; \quad \Gamma = \begin{pmatrix} 2\Gamma_R & -2\Gamma_B \\ -\Gamma_R & \Gamma_\omega \end{pmatrix} \qquad (17)$$

where we have taken $\Gamma_\omega = \Gamma_B + \Gamma_{ES}$ and $\Gamma_R = RN^0/Vol$ as the steady-state recombination rate. The eigenvalues of $\Gamma$ determine the time constants of the system's response to small perturbations. Gray showed that the dominant time constant for the quasiparticle response in the limit $\Gamma_R \ll \Gamma_B + \Gamma_{ES}$ is

$$\Gamma_R^* = 2\Gamma_R F_\omega^{-1} \quad ; \quad F_\omega = 1 + \frac{\Gamma_B}{\Gamma_{ES}} \tag{18}$$

where $F_\omega$ is called the phonon trapping factor. It accounts for a phonon emitted by a recombination event breaking another pair before it escapes to the bath. We note that $F_\omega^{-1}$ is just the probability that a phonon escapes to the bath. $\Gamma_R^*$ is the time constant with which a small perturbation of the quasiparticle system will decay, and it is the rate we expect to measure in experiments. We see that the measured recombination rate, $\Gamma_R^*$, is generally very different from the true equilibrium recombination rate $\Gamma_R$.

We can now specialize the multivariable master equation to describe the fluctuations in our intrinsic system. Again, we will describe the superconductor as a three level system consisting of quasiparticles, phonons in the superconductor and phonons in the bath. The three levels are connected by various transitions labeled $\{p_{ij}\}$ in Fig. 2. Each transition represents a physical process that changes the occupation of the three levels. Transition $p_{12}$ describes two quasiparticles recombining to create one phonon in the electrode. Transition $p_{21}$ describes the reverse process, a phonon being absorbed and generating two quasiparticles. Transition $p_{23}$ describes a phonon escaping from the electrode into the bath. Finally, $p_{32}$ describes a phonon entering the electrode from the bath. We note that there is no direct connection between levels 1 and 3, the quasiparticles and the bath. Since we have a three level system, our underlying master equation is a two variable equation. We choose as our two variables the number of quasiparticles, N, and the number of phonons in the electrodes, $N_\omega$. Referring to the rate equations for the system, (14) – (16), we can read off the transition probabilities, which we tabulate in Table I.

| Transition | Symbol | Probability per unit time |
|---|---|---|
| recombination | $p_{12}$ | $(1/2)RN^2/Vol$ |
| generation | $p_{21}$ | $\Gamma_B N_\omega$ |
| phonon escape | $p_{23}$ | $\Gamma_{ES} N_\omega$ |
| phonon entry | $p_{32}$ | $\Gamma_{ES} N_\omega^0$ |

**TABLE I. Allowed transitions and the probability per unit time for each one.**

In addition to the transition probabilities, we can also read off the shot size for each level, which is $\delta n_1 = 2$ for the quasiparticles and $\delta n_2 = 1$ for the phonons. Plugging these parameters into the above equations we find

$$\mathbf{M} = \begin{pmatrix} 2\Gamma_R & -2\Gamma_B \\ -\Gamma_R & \Gamma_\omega \end{pmatrix} \quad ; \quad \mathbf{B} = \Gamma_R N^0 \begin{pmatrix} 4 & -2 \\ -2 & 1 + \frac{\Gamma_{ES}}{\Gamma_B} \end{pmatrix} \qquad (19)$$

where $\Gamma_\omega = \Gamma_{ES} + \Gamma_B$. With these matrices we can then write the covariance matrix for our system. We find

$$\sigma^2 = \begin{pmatrix} N^0 & 0 \\ 0 & N_\omega^0 \end{pmatrix} = \begin{pmatrix} N^0 & 0 \\ 0 & \frac{1}{2}\frac{\Gamma_R}{\Gamma_B}N^0 \end{pmatrix}$$

where we have used the principle of detailed balance to relate $N_\omega^0$ to $N^0$. Thus, we again find that the variance of the occupation of each level is equal to the average occupation, as we expect from basic thermodynamic arguments. We also note that the off-diagonal terms are identically zero, implying that the quasiparticle and phonon fluctuations are independent. This is somewhat surprising since, as we will see later, the presence of the phonons does significantly modify the spectrum of the quasiparticle fluctuations.

Experimentally, we can only measure the spectrum of the quasiparticle fluctuations, so we will only calculate that spectrum. Using (13) and quite a bit of algebra, we obtain the quasiparticle spectrum

$$S(\omega) \equiv G_{11}(\omega) = \frac{2\alpha_1 \tau_1 N^0}{1+(\omega\tau_1)^2} + \frac{2\alpha_2 \tau_2 N^0}{1+(\omega\tau_2)^2} \qquad (20)$$

where

$$\alpha_1 = 2\frac{\tau_1 - \tau_{ES}}{\tau_1 - \tau_2} \quad ; \quad \alpha_2 = 2\frac{\tau_{ES} - \tau_2}{\tau_1 - \tau_2}$$

and $\gamma_{1,2} = 1/\tau_{1,2}$ are the eigenvalues of **M** and $\tau_{ES} = 1/\Gamma_{ES}$. It is straightforward to show that if we integrate $S(\omega)$ over all $\omega$ we recover $N^0$ for the variance. This expression is completely general. However, in the limit $\Gamma_R \ll \Gamma_B + \Gamma_{ES}$ we can simplify the eigenvalues of **M** to $\tau_1 = 1/\Gamma_R^*$ and $\tau_2 = (\Gamma_{ES} + \Gamma_B)^{-1}$, where $\Gamma_R^*$ is defined in (18). In this case, one time constant basically corresponds to the effective quasiparticle lifetime and one corresponds to the phonon lifetime. We can then interpret the first term of (20) as "intrinsic" quasiparticle fluctuations and the second term as phonon-driven fluctuations.

In a previous Letter, we presented experimental verification of these results by measuring quasiparticle number fluctuations in an Al box.[5] The box was formed by a volume, $Vol = 100$ μm$^3$, of thin-film superconducting Al. Two sides of the box were contacted by superconducting Ta leads. The Ta leads allow electrical contact to the box through the Cooper pair system, while still confining quasiparticles in the Al. Thermal quasiparticles in the Al cannot enter the Ta because the energy difference between the superconducting energy gap of Ta ($\Delta_{Ta} = 700 \mu eV$) and the energy gap of Al ($\Delta_{Al} = 180 \mu eV$) is much greater than $k_B T \approx 20 - 30 \mu eV$ and confines the quasiparticles. There are no thermal quasiparticles in the Ta at the temperatures used. We measured the number of quasiparticles by dividing the box with a tunnel barrier and measuring the current through the tunnel barrier. At large bias, the tunneling current is simply proportional to the number of quasiparticles in the box. By measuring the current noise spectrum of the biased junction, we were able to measure the spectrum of the quasiparticle fluctuations.

We also directly measured the recombination time of quasiparticles in the box with single-photon absorption experiments.[12] A single photon from the mercury emission line at 4.89 eV (256 nm) was absorbed in one Ta lead, producing about 4000 quasiparticles. These quasiparticles diffuse to the Al where they can emit phonons and drop down in energy, becoming trapped. These trapped quasiparticles are a small perturbation to the $N_0 \sim 10^5$ steady-state quasiparticles in the Al box. The trapped quasiparticles circulate, tunneling and backtunneling, until they are lost to recombination with a thermal quasiparticle. This produces a current pulse that decays exponentially on a time scale of the effective recombination time, $\tau_R{}^*$.

Now, in thin-film Al electrodes at the temperatures used we expect $\Gamma_R \approx 10^4 s^{-1}$, $\Gamma_{ES} \approx 10^9 s^{-1}$ and $\Gamma_B \approx 10^{10} s^{-1}$ (Ref. 13). Thus, we expect $\alpha_1 \approx 2(1 - 10^{-5})$ and $\alpha_2 \approx 2(10^{-5})$. This gives us a simplified expression for the spectrum:

$$S(\omega) \approx \frac{4 \tau_R{}^* N^0}{1 + (\omega \tau_R{}^*)^2}.$$

Our measurements showed good agreement with this result. First, we confirmed that the quasiparticle fluctuations had this Lorentzian form. We also confirmed that the characteristic time of the fluctuations was in fact $\tau_R^*$, over a range of temperatures by comparing the noise measurements to the direct measurements of $\tau_R^*$. Finally, we demonstrated that magnitude of the noise agreed with our prediction over the same temperature range.

If we compare this simplified $S(\omega)$ with the one-variable result found in (7), we see that this power spectrum could have been obtained from a simpler one-variable master equation assuming effective generation and recombination parameters

$$r(N) = \frac{1}{2} \frac{R}{F_\omega Vol} N^2 \quad ; \quad g(N) = r(N^0)$$

where the generation parameter g(N) is just a constant equal to the equilibrium recombination rate. This simplification is not general, but it is possible in our samples because the quasiparticle and phonon time scales are so widely separated. Basically, the quasiparticle system cannot respond to the fast phonon fluctuations and is only affected by the average number of phonons.

*C. Extrinsic quasiparticle fluctuations*

For our second case, we consider extrinsic quasiparticle fluctuations where we allow quasiparticles to be lost to processes other than recombination. In particular, we consider additional processes that change the number of quasiparticles by one. There are many physical examples of this kind of process including trapping into material defects,[14] diffusion, trapping into normal-metal regions induced by fluxons,[15] and trapping into external normal-metal "sinks."[16] The multivariable theory presented here could be applied to a system with an arbitrary number of these extrinsic loss processes. However, we will develop the theory for only one extrinsic loss process in addition to intrinsic recombination. If we were to consider such a system fully, including phonons, we would have a four level system described by a three variable master equation. However, we saw in the previous section that in many relevant experimental systems the effect of the phonons reduces to simply modifying the recombination constant. We therefore consider only a three level system with an effective recombination constant $R^*$.

Our three levels are: 1) the number of free quasiparticles, 2) the number of trapped quasiparticles, and 3) the number of pairs. The levels are described by the occupation numbers $N$, $N_t$, and $N_p$ respectively, and we take $N$ and $N_t$ to be independent. We assume the allowed transition parameters are $p_{12} = \Gamma_t N$, $p_{21} = \Gamma_d N_t$, $p_{13} = R^* N^2/(2Vol)$, and $p_{31} = p_{13}^0 = R^*(N^0)^2/(2Vol)$ where $\Gamma_t$ is the trapping rate and $\Gamma_d$ is the detrapping rate. We also write the shot sizes as $\delta n_{12} = 1$, $\delta n_{21} = 1$, $\delta n_{13} = 2$, and $\delta n_{31} = 1$. We have made some implicit assumptions in writing these transition parameters. First, we have assumed that we are working at low temperatures such that the number of pairs is much greater than the number of quasiparticles. Second, we have assumed that our traps are far from being saturated, such that the transition parameters do not depend on the number of available trap states. With these parameters and assumptions we can apply equations (9) and (10) to find

$$\mathbf{M} = \begin{pmatrix} \Gamma_R{}^* + \Gamma_t & -\Gamma_d \\ -\Gamma_t & \Gamma_d \end{pmatrix} ;$$

$$\mathbf{B} = \begin{pmatrix} 2(\Gamma_R{}^* + \Gamma_t)N^0 & -(\Gamma_t N^0 + \Gamma_d N_t^0) \\ -(\Gamma_t N^0 + \Gamma_d N_t^0) & 2\Gamma_d N_t^0 \end{pmatrix} = 2N^0 \begin{pmatrix} \Gamma_R{}^* + \Gamma_t & -\Gamma_t \\ -\Gamma_t & \Gamma_t \end{pmatrix}$$

where $\Gamma_R{}^* = 2R^*N^0/Vol$. In simplifying $\mathbf{B}$, we have applied the principle of detailed balance, i.e., assumed $p_{21}{}^0 = p_{12}{}^0$ and $p_{31}{}^0 = p_{13}{}^0$. As discussed earlier, this is always valid for a system in thermodynamic equilibrium, but it must also be true for our system in steady-state or $N_t$ and $N_p$ would not have well-defined steady-state values. We can therefore use the quasiequilibrium result (11) to calculate the covariance matrix

$$\sigma^2 = \begin{pmatrix} N^0 & 0 \\ 0 & N_t^0 \end{pmatrix} = \begin{pmatrix} N^0 & 0 \\ 0 & \frac{\Gamma_t}{\Gamma_d} N^0 \end{pmatrix}.$$

We see that, even though the quasiparticles are now connected to more than one level, the variance of their fluctuation is still simply $N^0$.

We could now calculate the general power spectra of this model, but the equations are not particularly illuminating. Instead, we will further simplify the model to the experimentally interesting case where trapping and detrapping are the faster processes. Specifically, we will assume that $\Gamma_t + \Gamma_d \gg \Gamma_R{}^*$. In this limit, the eigenvalues of $\mathbf{M}$ are:

$$\gamma_1 = \frac{\Gamma_d}{\Gamma_d + \Gamma_t} \Gamma_R{}^* \; ; \; \gamma_2 = \Gamma_d + \Gamma_t.$$

The spectrum of the quasiparticle fluctuations is then

$$S(\omega) \equiv G_{11}(\omega) = \frac{S_1}{1 + (\omega \tau_1)^2} + \frac{S_2}{1 + (\omega \tau_2)^2} \qquad (21)$$

where

$$S_1 = 4N^0 \tau_1 \left[ \frac{\Gamma_d - \Gamma_R{}^*}{\gamma_2 - \gamma_1} \right] \approx 4N^0 \tau_1 \frac{\Gamma_d}{\Gamma_d + \Gamma_t}$$

and

$$S_2 = 4N^0\Gamma_t(\tau_2)^2\left[\frac{\gamma_2+\Gamma_R{}^*}{\gamma_2-\gamma_1}\right] \approx 4N^0\tau_2\frac{\Gamma_t}{\Gamma_d+\Gamma_t}$$

where the final simplification of $S_1$ and $S_2$ represent extreme limits. As before, the spectrum is the sum of two Lorentzians each with a bandwidth determined by the eigenvalues of **M**. The relative weight of each Lorentzian depends on the depth of the traps. We call the traps "deep" if $\Gamma_t > \Gamma_d$, meaning that once a quasiparticle is trapped it takes a relatively long time for it to escape. Conversely, we call the traps "shallow" if $\Gamma_d > \Gamma_t$, meaning that quasiparticles escape relatively quickly. For very deep traps $\gamma_2 \approx \Gamma_t$ and $S_2$ dominates $S_1$, such that

$$S_{deep}(\omega) \approx \frac{4\tau_t N^0}{1+(\omega\tau_t)^2}$$

where $\tau_t = 1/\Gamma_t$. This is the result we would expect for a two-level system where quasiparticles can be lost only to traps. For very shallow traps, we instead find

$$S_{shallow}(\omega) \approx \frac{4\tau_R{}^* N^0}{1+(\omega\tau_R{}^*)^2}$$

which is the result we expect for quasiparticles in the presence of recombination only. Eqn. (21) varies smoothly between these two cases and it is easy to show that the integral of the power spectrum is $N^0$ for any trap depth.

## IV. DISCUSSION

### A. Theoretical connections

We note that comparing the two rate matrices (17) and (19), $\mathbf{\Gamma} = \mathbf{M}$ for the intrinsic fluctuation problem. As we have seen, the eigenvalues of $\mathbf{M}$ determine the spectrum of the fluctuations. On the other hand, the eigenvalues of $\mathbf{\Gamma}$ determine the time constants of the dynamical response to small perturbations. The fact that these two matrices are equal implies that the timescales of the dynamical response and the fluctuations are the same. We will now take some time to explore the generality of this connection beyond the specific example of intrinsic fluctuations.

When we write rate equations like (14) – (16) we are making some implicit approximations. First, we approximate the occupation numbers of the levels, such as N and $N_\omega$, as continuous variables, when they are in fact discrete variables. Second, we approximate the discrete and random transitions between levels as continuous and deterministic "flows". To understand the implications of these approximations, we start by deriving a differential equation for the expectation values of the level occupations from the master equation (8). We find the following system of equations:

$$\frac{\partial}{\partial t}\langle N_i \rangle = \sum_{j \neq i} \delta n_i \left( \langle p_{ji}(N_j) \rangle - \langle p_{ij}(N_i) \rangle \right) \tag{22}$$

where the indices i and j run over all levels. We can compare this equation to a general expression for the rate equations (similar to (6)), which is

$$\frac{\partial}{\partial t} N_i = \sum_j \delta n_i \left( p_{ji}(N_j) - p_{ij}(N_i) \right). \tag{23}$$

The only apparent difference is that we have dropped the expectation value brackets from the second system of equations. However, we must also keep in mind the subtle difference that the first equation is an exact differential equation for the continuous expectation value of a discrete variable. The second equation is only approximate, for the reasons mentioned above.

However, in the special case where the $\{p_{ij}(N_i)\}$ are all linear functions of the occupation numbers, $\{N_i\}$, we have that $\langle p_{ij}(N_i)\rangle = p_{ij}(\langle N_i\rangle)$ and we can actually interpret the rate equations (23) as exact equations for the expectation values of the occupations of the levels. In many physical systems, although, the transition rates are at least quadratic in the occupation numbers, such that, $p_{ij} \sim N_i^2$ or $p_{ij} \sim N_iN_j$. In this case, we have, for example, that

$$\langle p_{ij}\rangle \sim \langle N_i^2\rangle = \langle N_i\rangle^2 + \langle \Delta N_i^2\rangle.$$

Thus, in the case of quadratic transition rates we can interpret the rate equations as approximate equations for the expectation values, ignoring terms of order the variance of the occupation number. In general though, we expect that $\langle \Delta N_i^2\rangle \sim \langle N_i\rangle$ and we can say that neglecting the variance terms is a valid approximation to order $\mathcal{O}(1/N)$. In other words, for a large system the rate equations actually describe the expectation values of the occupation numbers, to good approximation.

We can develop this idea a little further. If we take (22) and Taylor expand the transition probabilities to first order we get the following equation for small variations:

$$\frac{\partial}{\partial t}\langle \Delta N_i\rangle = \sum_{j\neq i}\delta n_i\left[\sum_k \frac{\partial p_{ji}}{\partial N_k}\langle \Delta N_k\rangle - \sum_k \frac{\partial p_{ij}}{\partial N_k}\langle \Delta N_k\rangle\right]_{\{N_k\}=\{N_k^0\}} = \sum_k M_{ik}\langle \Delta N_k\rangle$$

where $M_{ik}$ are the elements of the matrix **M** defined in (9). If we follow the same procedure for the rate equations, and we find that the linearized rate equations:

$$\frac{\partial}{\partial t}\Delta N_i = \sum_{j\neq i}\delta n_i \left[\sum_k \frac{\partial p_{ji}}{\partial N_k}\Delta N_k - \sum_k \frac{\partial p_{ij}}{\partial N_k}\Delta N_k\right]_{\{N_k\}=\{N_k^0\}} \equiv \sum_k \Gamma_{ik}\Delta N_k$$

where have defined the linearized rate matrix $\Gamma$. We see, in general now, that $\Gamma = \mathbf{M}$ and that we can interpret the linearized rate equations as equations for the expectation values around equilibrium. This is the general connection between the fluctuations and the dynamics. The eigenvalues of $\Gamma$ determine the dynamical response of the system to perturbations. The eigenvalues of $\mathbf{M}$, in turn, determine the characteristic times of the fluctuations. But, as we have just seen, the matrices are the same. Thus, the timescales measured from dynamic perturbations and from equilibrium fluctuations must be the same. The proceeding discussion amounts to a statistical fluctuation-dissipation theorem for our system. In fact, we can derive the fluctuations of our system in a more conventional thermodynamic framework using the fluctuation-dissipation theorem.[7,17]

## B. Experimental Connections

The general connection between dynamics and fluctuations implies that measurements of fluctuations in a system are a useful probe of dynamical timescales. A common way to probe the dynamics of a system is perturb the system slightly and then observe the decay of the perturbation. If fluctuations of the system can be observed, however, the perturbation method is no longer necessary. Indeed, for a variety of reasons, it may be preferable to avoid perturbing the system. For example, we must always wonder to what extent the relaxation of an external perturbation represents the intrinsic behavior of the system. In addition, providing a perturbation to the system, especially a small, controlled perturbation, may be experimentally challenging.

In superconducting systems, at least in principle, the fundamental time scale of electron-phonon interactions, known as $\tau_o$, can be inferred from measurements of quasiparticle-quasiparticle recombination. The parameter $\tau_o$ is material dependent and is of general interest because it sets the timescale of many processes related to the interaction of electrons and phonons.[13] In particular, for a pair of quasiparticles at the gap edge, the expression for the recombination constant is

$$R = \left(\frac{2\Delta}{k_B T_c}\right)^3 \frac{1}{2\Delta D(\varepsilon_F)\tau_o} \qquad (24)$$

where $\Delta$ is the superconducting energy gap, $T_c$ is the superconducting transition temperature and $D(\varepsilon_F)$ is the electron density of states at the Fermi energy. However, phonon trapping complicates the extraction of $\tau_o$ from recombination measurements at temperatures much less than $T_c$. In fact, in the limit of strong phonon trapping, the measured recombination rate $\Gamma_R^*$ becomes

$$\Gamma_R^* = 2\frac{\Gamma_R}{\Gamma_B}\Gamma_{ES} \sim \Gamma_{ES}$$

because the pair-breaking rate, $\Gamma_B$, is also proportional to $1/\tau_o$. Thus, measurements of $\Gamma_R^*$ in the presence of strong phonon trapping have no dependence on $\tau_o$.

As described above, we have used both fluctuations and photoexcitation to measure $\Gamma_R^*$ in Al. If we ignore phonon trapping for the moment and insert our measured value of R into (24), we extract a tentative value for $\tau_o$ of 1.65 μs. Numerous other measurements of $\tau_o$ in Al by various methods find values of order 100 ns.[11,18] This discrepancy suggests that our measurements are, in fact, in the limit of strong phonon trapping, so that they do not represent a direct measurement of $\tau_o$. Our measurements do, however, confirm that the quasiparticle recombination rate is proportional to the quasiparticle density at lower temperatures and longer recombination times than previous experiments. In Fig. 3, we compare the recombination time measured by us to previous experiments and to theory. The previous measurements showed recombination times that begin to deviate from the expected dependence at $T \approx 400\ mK$ and $\tau_R^* \approx 20\ \mu s$ and completely saturate at a maximum value of $\tau_R^* \approx 80\ \mu s$ below $T \approx 300\ mK$.[19] Quasiparticle loss into normal-metal regions created by trapped flux was proposed as the explanation for the deviation from theory in those measurements, although this explanation was not experimentally confirmed. Therefore, our measurements extend the range over which the basic physics of recombination has been verified in Al.

While quasiparticle number fluctuations may be helpful in studying the microscopic dynamics of superconductors, they are also a source of noise in superconducting electronic devices. We have discussed in detail how they can limit the performance of single-photon spectrometers based on superconducting tunnel junctions.[20] In addition, quasiparticle fluctuations may be an important source of noise, and therefore decoherence, in superconducting quantum bits (qubits). The majority of solid-state systems that have been used to demonstrate coherent quantum manipulation of a single qubit have involved superconductors.[3] All of these measurements have been performed at very low temperatures ($T/T_c \sim 0.01$), where there would be essentially zero quasiparticles in equilibrium. However, all of the readout schemes in these experiments produce nonequilibrium quasiparticles, which can accumulate in the qubits, leading to a steady-state density of quasiparticles. At least one experiment has directly demonstrated the impact of this quasiparticle background on the measured coherence times.[21] Understanding the effect of quasiparticle fluctuations on coherence may therefore be important for the development of quantum bits.

In conclusion, we have developed a general theory of quasiparticle number fluctuations in superconductors. We applied this general theory to the problem of intrinsic quasiparticle fluctuations related to generation and recombination. The validity of these results have been demonstrated in previous experimental work. We have also applied the theory to an example of extrinsic quasiparticle fluctuations where quasiparticles are also lost to traps. We conclude that studies of quasiparticle fluctuations provide a useful probe of microscopic dynamics and are also important for the understanding of noise in superconducting devices.

## ACKNOWLEDGEMENTS

We would like to thank Luigi Frunzio, Michel Devoret, Robert Schoelkopf, and Liqun Li for help and useful discussions. Funding for this work was provided by NASA-NAG5-5255 and NSF-DMR-0072722. The first author was supported in part by a NASA GSRP fellowship and the Keck Foundation.

* Electronic address: christopher.wilson@yale.edu and daniel.prober@yale.edu

[1] For example: A. W. Kleinsasser, IEEE Trans. Appl. Supercond. **11**, 1043 (2001); P. Bunyk, K. Likharev and D. Zinoviev, Int. J. of High Speed Elec. and Sys. **11**, 257 (2001).

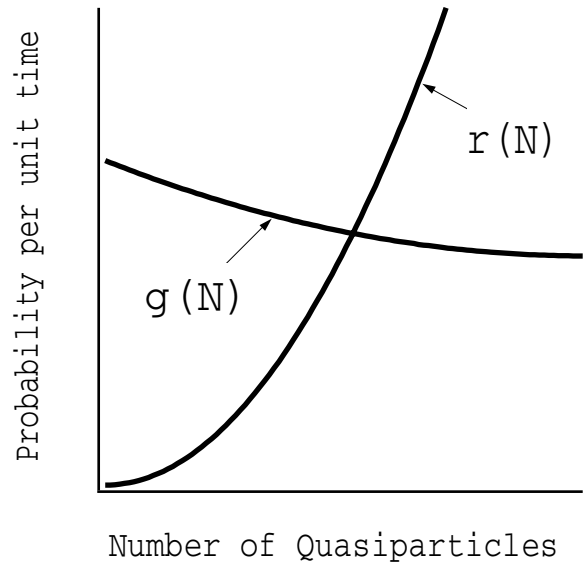

**FIG. 1. Sketch of the generation and recombination parameters, g(N) and r(N). The intersection of the curves yields the steady-state number of quasiparticles.**

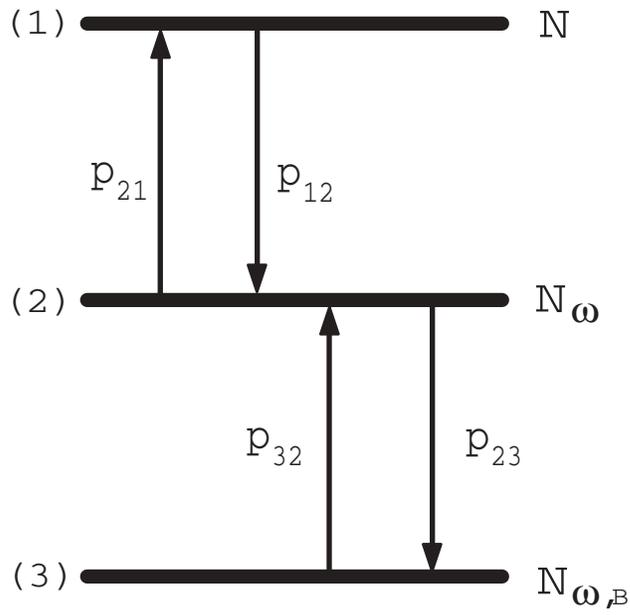

**FIG. 2. Schematic representation of our three level system. From top to bottom, the levels are quasiparticles, phonons in the electrodes, and phonons in the bath.**

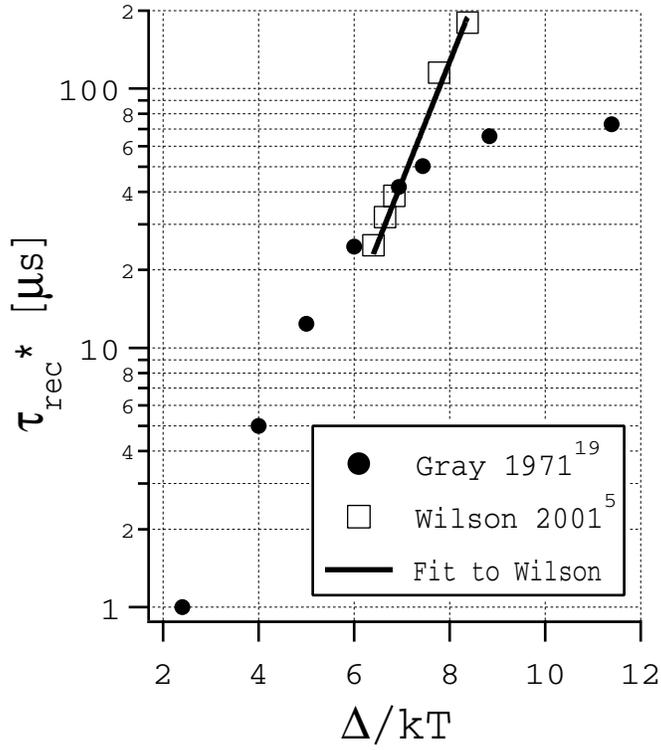

FIG. 3. Comparison of quasiparticle lifetime measurements described here to previous measurements by Gray.[19] Measurements by Gray were on Al on sapphire with $\Delta$ = 195 µV. Our films are on $SiO_2$ with $\Delta$ = 180 µV. The solid line shows the theoretical scaling of the lifetime with the BCS number of quasiparticles for our value of $\Delta$. Our data show the lifetime following the theoretical dependence to lower temperature.